\documentstyle{article}
\newcommand{\arcm}{\hbox{$^\prime$}}
\newcommand{\arcsec}{\arcm\hskip-0.1em\arcm}
\newcommand{\uarcs}{\hskip-0.27em\arcsec\hskip-0.02em}
\begin{document}

\normalsize

\begin{center}
\vspace{12mm}
{\large STUDIES OF THE VIRGO CLUSTER BASED ON\\ THE VIRGO PHOTOMETRY CATALOGUE\\}
\vspace{3mm}
{\small {\large C}HRISTOPHER {\large K}E-SHIH {\large Y}OUNG}\\
{\normalsize Beijing Astronomical Observatory, Chinese Academy of Sciences, Beijing 100080\\}
\vspace{5mm}
{\normalsize ABSTRACT}
\vspace{-3mm}
\end{center}

\noindent 
The {\em Virgo Photometry Catalogue\/} (VPC) is the first 
independently calibrated
general catalogue of galaxies to cover the Virgo cluster since the {\em Catalog of 
Galaxies and Clusters of Galaxies\/} of Zwicky et al. (1961,1963). 
It contains 1180 galaxies (including background objects) within a 23 square-degree 
region centred on the cluster's core.
Photographic surface photometry is presented for 1067 galaxies in the $U$ band, 
for 1020 galaxies in the $B_J$ band and for 1020 galaxies in the $R_C$ band.
All total magnitudes and total colours are extrapolated according to a new system;
denoted $t$ to distinguish it from the $T$ system already in use.
This paper outlines: the scope of the VPC, the new extrapolation system,
some recent findings based on the catalogue and further work in progress.

\vspace{1mm}

\begin{center}
1. THE CATALOGUE
\vspace{-3mm}
\end{center}

The VPC is based primarily on four U.K.-Schmidt plates (one $U$, two $B_J$ and one $R_C$)
which were digitised using the Royal Observatory Edinburgh's {\em COSMOS\/} two-dimensional
scanning microdensitometer. The centre of the VPC field is  
R.A.$_{1950.0}=12^{\rm{h}}26^{\rm{m}}$ and Dec.$_{1950.0}=13^{\circ}08$\arcm.
The galaxy sample consists of non-stellar objects brighter than
$B_{J_{25}}=19.0$; the completeness limits being
$B_{J_{25}}\approx18.5$ for the northern half of the survey area and
$B_{J_{25}}\approx18.0$ for the southern half.
Parameters listed for catalogued galaxies include:
equatorial coordinates, morphological types, surface-brightness profile parameters (as well as
the original surface-brightness profiles), $U$, $B_J$ \& $R_C$ isophotal 
and $t$-system total magnitudes, transformed $B_t$ magnitudes, $(U-B_{J})$ \& $(B_{J}-R_{C})$ 
equal-area and $t$-system total
colours, apparent angular radii, ellipticities, position angles, heliocentric radial velocities,
alternative designations and cluster-membership assignments. All magnitudes and
colours were derived from numerical integrations of segmented plate-scan data, except for
(in 109 cases) saturated or (in 51 cases) inextricably merged images; the segmentation
software having been able to cope with the vast majority of image mergers.

The $U$-band photometry was calibrated by means of 280 $U$-band aperture-photometry measurements from
the literature; whilst the $R_C$-band photometry was calibrated by means of 16 published $R_C$-band 
aperture-photometry measurements, as well as four $R_C$-band CCD 
frames.
The $B_J$-band photometry was calibrated by means of 239 pairs of $B$- and 
$V$-band aperture-photometry measurements from the literature and the transformation: $B_{J}= 
B-\alpha(B-V)$, with $\alpha=0.35$. However,
were the [now more
fashionable] value of $\alpha=0.28$ to be more accurate, the VPC's $B_J$ photometric zero point 
may be too faint by of the order of 0.07 mag.

The author would therefore not wish to pretend that the zero-points of the VPC's $B_J$ and
transformed $B$ magnitude scales are of the highest accuracy attainable. However, from the
comparisons between the VPC's surface photometry and well-documented photoelectric aperture
and CCD surface photometry in the literature, as described by Young (1994) and
Young \& Currie (submitted), it appears very unlikely that the VPC's $B_J$ and/or
transformed $B$ zero points could be 
too faint by more than 0.1 mag., whilst it is even more unlikely that they could be
even marginally too bright.

An early version of the catalogue has already been presented by Young (1994) but the final
version is presented by Young \& Currie (submitted), in which the minor differences between the 
two versions of the catalogue are quantified. The main difference between the two 
versions is that in the final version, the ${B_J}_t$ values are derived directly from Equation~3 
rather than by a slightly less rigorous numerical method. 
The final $B_{J_t}$ values are on average marginally brighter by 0.004 mag.
Work already based on the early version of the catalogue includes:
Young \& Currie (1995) (whose subsample of 64 VPC dwarf ellipticals
are on average 0.026 mag.\ brighter in the final catalogue)
and Drinkwater et al.\ (1996).

\newpage

\begin{center}
2. THE $t$ SYSTEM FOR ESTIMATING TOTAL MAGNITUDES
\vspace{-3mm}
\end{center}

This new system differs fundamentally from the $T$ system of de Vaucouleurs et al.'s (1976, 1991) {\em Second\/} and
{\em Third Reference Catalog of Bright Galaxies\/} (RC2, RC3) and 
of de Vaucouleurs \& Pence (1979); in that (1) 
extrapolations of photometric profiles do {\em not\/} rely on any prior morphological classification and (2) the smearing 
effects that degrade the resolutions of photometric profiles are {\em not\/} neglected. 
In this system, two-dimensional digital galaxy images are [unless already of 
sufficiently low resolution] first smoothed numerically (ideally convolved with a 
radially symmetric Gaussian function) in order to reduce their resolution 
sufficiently that their
surface-brightness profiles can be modeled by S\'{e}rsic's (1968) 
law irrespective of galaxy morphological type:

\begin{equation}
\sigma(r)= \mbox{ } \sigma_{0} \mbox{ } \exp [-(\frac{r}{r_{0}})^{n}],
\end{equation}

\noindent
in which $r$ is the reduced radial distance ($\sqrt{r_{\rm major}r_{\rm minor}}$),
$\sigma(r)$ is the surface brightness in linear units of luminous flux density at 
$r$,
$\sigma_0$ is the central surface brightness and $r_0$ is the angular scale-length.
In standard logarithmic surface-brightness units, the central surface 
brightness is therefore $\mu_{0}=-2.5 \log_{10}\sigma_{0}$ in mag.arcsec$^{-2}$, 
whence the equivalent expression:

\begin{equation}
\mu(r)= \mbox{ } \mu_{0} \mbox{ } + \mbox{ } 1.086[-(\frac{r}{r_{0}})^{n}],
\end{equation}

\noindent
enabling values of $\mu_0$ and $r_0$ to be obtained by linear regression when the optimum
value of $n$ has been derived. The analytical solution:

\begin{equation}
2\pi\int_{0}^{\infty} \sigma_{0} r \mbox{ } \exp [-(\frac{r}{r_0})^{n}] \mbox{ } dr=
\mbox{ } \frac{2}{n} \pi \sigma_{0} \Gamma(\frac{2}{n}) {r_{0}}^{2},
\end{equation}

\noindent
then yields an estimate of the total luminous flux within the pass-band concerned.

Clearly, this generalisation incorporates not only the $r^{\frac{1}{4}}$ law
($n=\frac{1}{4}$) but also exponentials ($n=1$) and Gaussians ($n=2$) as well
as both intermediate and more extreme cases, though it is still a one-component
model. 
As is evident from Young \& Currie (1994, 1995) images of 
dwarf-elliptical galaxies whose surface-brightness profiles can be fitted well by a single component 
model do not generally need to be smoothed except when they are nucleated.

This extrapolation system has already been outlined by Young \& Currie (1994, 1995). It is described 
in more detail by Young et al.\ (submitted).

\vspace{1mm}

\begin{center}
3. ERRORS IN EXISTING MAGNITUDE SCALES
\vspace{-3mm}
\end{center}

A magnitude scale was originally established for Virgo galaxies by
Zwicky et al.\ (1961) and Zwicky \& Herzog (1963) in Volumes I and II respectively
of their {\em Catalog of Galaxies and Clusters of Galaxies\/} (CGCG).
de Vaucouleurs \& Pence (1979) later applied transformations to the CGCG magnitudes
of objects in the direction of the Virgo cluster
(as well as to magnitude values quoted in two other less-extensive catalogues
for these CGCG galaxies) with a view to reducing them to 
their blue-band total-magnitude ($B_T$) scale based on their 
$T$-system extrapolations.

These transformed magnitudes were then used by
Binggeli et al.\ (1984) for the zero-point calibration of Binggeli et al.'s (1985) blue-band 
{\em Virgo Cluster
Catalog\/} (VCC), which went much deeper than either the CGCG or
de Vaucouleurs \& Pence's catalogue. As the VCC is the most extensive catalogue
of Virgo galaxies published to date, it has become a standard reference work and
the main source of magnitudes for Virgo galaxies, even though it lacks any independent zero-point 
calibration.

For the brighter Virgo galaxies, another frequently used source of magnitude
measurements has long been the {\em Reference Catalog of Bright Galaxies\/}
series. The latest edition in this series is the RC3.
This catalogue is primarily a compilation of magnitudes from different sources
transformed to a common system, but its magnitude values for Virgo galaxies
were not independently calibrated.

Comparisons between the magnitude scales of the VPC and those of existing studies of Virgo galaxies 
(as well as between the existing studies themselves), suggest serious systematic errors throughout 
most of the literature. Young (1994) has already found that the transformations of de Vaucouleurs 
\& Pence (1979) appear to suffer from very serious systematic errors at the faint end, that lead to 
over-estimates of galaxy luminosities by $^{<}_{\approx}$0.7 mag.\ for their Table~2 objects and 
$^{<}_{\approx}$1.1 mag.\ for their 
Table~4 objects. The transformations used in the compilation of the RC2 and by de 
Vaucouleurs \& Pence were eventually renounced, and thereby 
superseded, in the RC3 (1991). However, in the meantime, Binggeli et al.\ (1984) based the VCC's 
magnitude scale on the original work of de Vaucouleurs \& Pence. Young (1994) found that the VCC appears to
overestimate galaxy luminosities by $^{<}_{\approx}$0.7 mag.\ at $B_{t}=16.5$ (or their $B_T\approx15.8$).

The RC3 transformations, whilst a significant improvement on those of de Vaucouleurs \& Pence
still appear to lead to significant over-estimates of galaxy luminosities faintward of
$B_{t}\sim15$. The mean systematic error may still be as much as 
$\sim$0.5 mag.\
at $B_{t}=16.0$. It can be attributed in large part to the application of inappropriate (e.g.\ 
$r^{\frac{1}{4}}$-law) extrapolations to low surface-brightness galaxies 
(such as IC~3475 which Vigroux et al.\ 1986 have already demonstrated
to have an exponential profile)
during the compilation of the RC3, as discussed in detail by Young et al.\ (submitted). 
Therefore, the RC3 magnitude scale may well suffer from similar systematic errors in other 
fields  with high surface number densities of low-surface-brightness galaxies, such as the direction of the Fornax cluster.

The main conclusion thus far is therefore that differences between the extrapolations of the 
pre-RC3 $T$ system and those of the $t$ system, are primarily responsible for the extremely large
systematic differences between the total magnitudes quoted in both the RC2 and the VCC on the one
hand, and those quoted in the VPC on the other hand. Disagreements between the zero-points of the
VPC's surface photometry and those of previously published photoelectric (including CCD) 
photometry cannot be the cause, because they are generally small. 
Saturation effects cannot be the cause either because 
care has been taken to flag all saturated image profiles
and in any case the observations were made under 
poor seeing (3.\uarcs0 and  1.\uarcs7 FWHM).
The example of IC~3475 can be used to illustrate the point. Vigroux et al.\ (1986) 
extrapolate its $B$-band profile to $r=\infty$ according to the exponential law and obtain a 
total magnitude of 14.47. In the RC3 on the other hand, aperture photometry for the same galaxy is extrapolated as 
if it obeyed the $r^{\frac{1}{4}}$ law yielding $B_T$=13.82. From Young (1994) and Young \& Currie
(submitted), differences between the $B$- \& $V$-band
aperture photometry used in the RC3 and simulated-aperture photometry 
based on unsmoothed VPC $B_J$-band plate-scan data are known to be negligible 
[by transforming between the $BV$ and $B_J$ systems]. Nevertheless, the VPC's $B_{J_t}$ value of 14.10 corresponds to a
transformed $B_t$ magnitude indistinguishable from Vigroux et al.'s value, and is thereby $\approx$0.7 mag.\ 
fainter than the RC3 value. 

The preliminary evidence for serious systematic errors in the faint ends of the existing magnitude
scales has already been presented by Young (1994), whose work is being supplemented by
$B$ and $V$-band wide-field CCD frames
obtained earlier this year (1996). These frames cover virtually the whole of the VPC
survey area and will yield 
independently calibrated $t$-system total magnitudes for the 109 brighter Virgo galaxies whose
photographic photometry was saturated in the VPC. A more detailed study  
quantifying the systematic errors as a function of magnitude (i.e.\ not just at the
faint end) will subsequently be presented by Young et al.\ (in preparation).

\vspace{1mm}

\begin{center}
4. THE VIRGO CLUSTER'S DISTANCE AND STRUCTURE
\vspace{-3mm}
\end{center}

Most attempts to measure the Hubble parameter have long been based on distance
measurements to Virgo-cluster galaxies. There are probably two main reasons for this. First, 
it is the nearest cluster in which a full complement of morphological types is represented.
It has therefore been possible to apply a wider variety of different distance indicators to
Virgo galaxies than it has been for other clusters at similar redshifts such as Fornax or Leo.
Second, Virgo has the added advantage of being easily observable from observatories in both
the Northern and Southern hemispheres.

Virgo's popularity as a pivotal step in the cosmic distance scale has nevertheless not been without 
its difficulties and controversies. Until recently, the possibility that the spatial extent of the
Virgo cluster may have considerable line-of-sight depth was not taken seriously. Although
Pierce \& Tully (1988) suggested the `possible presence of superposed foreground and background
galaxies', it was not until the distance-scale work of Tonry et al.\ (1990) and Tonry (1991)
that any author was prepared to stand by the case for considerable depth. However, these authors
did not confront the potentially embarrassing issue of the cluster's elongation ratio in the
line of sight, which would have to exceed  5:1 in order to explain the depth in their derived
galaxy distributions. Other authors have quite understandably tended to explain any apparent
depth effect in Virgo galaxy distributions in terms of random observational errors and scatter
intrinsic to the distance indicators employed.

The first authors {\em not\/} to avoid the elongation issue were Fukugita et al.\ (1993).
These authors suggested that the Virgo cluster's spiral galaxy distribution is in fact 
filamentary in shape, rather than approximately spherically symmetric [as had been assumed before],
with spirals lying at distances from 13 Mpc all the way to 30 Mpc. Young \& Currie (1995)
also found a considerable depth effect, that they could not explain away in terms of
intrinsic scatter in their $L$-$n$ and $R$-$n$ indicators; this time 
in the distribution of 64 VPC dwarf elliptical galaxies in the direction
of the Virgo cluster's core itself. They also found preliminary evidence for a foreground cluster
at a distance of about 9 Mpc, which led them to suggest that the depth effect may be due to
the projection of more than one distinct galaxy concentration in the same sight line. This
would avoid the need for elongation ratios of the order of 5:1. 

Further work is already in progress to enlarge the dwarf-elliptical galaxy sample 
(by extending the morphological-typing limit of the VPC to fainter magnitudes) 
so that a representative sample of tracer galaxies can be defined. This should enable the separation
of separate galaxy concentrations in the same sight line should they be present. This work
will be supplemented by as yet unpublished radial-velocity measurements for Virgo dwarfs, 
which will enable a kinematical study of the dwarfs.

\vspace{1mm}

\begin{center}
5. FURTHER STUDIES BASED ON THE VPC
\vspace{-3mm}
\end{center}

An investigation is already underway into the luminosity functions of Virgo dwarfs.
This new work on the subject will have the triple benefits over previous studies of:
galaxy samples selected according to objective criteria,
independently calibrated machine generated magnitude measurements and
the consideration of depth effects. Another study based on the VPC galaxy sample will be 
a pencil-beam survey in the Virgo direction. This study will be enhanced by new 
velocity measurements which will supplement those of Drinkwater et al.\ (1996).

\vspace{3mm}

\noindent
The author is grateful to the QSO \& Observational Cosmology Group at BAO
for use of its computing facilities. This research is currently supported by the 
National Postdoctoral Fellowship Office of China.

\vspace{1mm}

\begin{center}
REFERENCES
\vspace{-3mm}
\end{center}

\small

\noindent
Binggeli B., Sandage A., Tammann G.A., 1985, AJ, 90, 1681 (VCC)\\
Binggeli B., Sandage A., Tarenghi M., 1984, AJ, 89, 64 \\
de Vaucouleurs G., de Vaucouleurs A., Corwin H.G., 1976, Second Reference Catalog of Bright Galaxies, Uni-\\
\indent{versity of Texas, Austin (RC2)}\\
de Vaucouleurs G., de Vaucouleurs A., Corwin H.G., Buta R.J., Paturel G., Fouque P., 1991, Third Reference\\
\indent{Catalog of Bright Galaxies, Springer-Verlag, New York (RC3)}\\
de Vaucouleurs G., Pence W.D., 1979, ApJS, 40, 425\\
Drinkwater M.J., Currie M.J., Young C.K., Hardy E., Yearsley J.M., 1996, MNRAS, 279, 595\\
Fukugita M., Okamura S., Yasuda N., 1993, ApJ, 412, L13\\
Pierce M.J., Tully R.B., 1988, ApJ, 330, 579\\
S\'{e}rsic J.L., 1968, Atlas de galaxias australes, Observatorio Astronomica, Cordoba\\
Tonry J.L., 1991, ApJ, 373, L1\\
Tonry J.L., Ajhar E.A., Luppino G.A., 1990, AJ, 100, 1416\\
Vigroux L., Thuan T.X., Vader J.P., Lachieze-Rey M., 1986, AJ, 91, 70\\
Young C.K., 1994, D.Phil. Thesis, University of Oxford\\
Young C.K., Currie M.J., 1994, MNRAS, 268, L11\\
Young C.K., Currie M.J., 1995, MNRAS, 273, 1141\\
Zwicky F., Herzog E., 1963, Catalog of Galaxies and Clusters of Galaxies, 2, California Institute of Tech-\\
\indent{nology, Pasadena (CGCG)}\\
Zwicky F., Herzog E., Wild P., 1961, Catalog of Galaxies and Clusters of Galaxies, 1, California Institute\\
\indent{of Technology, Pasadena (CGCG)}
\end{document}